\documentclass{PoS}

\title{Japanese VLBI Network }

\ShortTitle{Japanese VLBI Network }

\author{\speaker{Akihiro~Doi}\footnote{E-mail: doi@yamaguchi-u.ac.jp},$^a$ 
Kenta~Fujisawa,$^a$ 
Keiichiro~Harada,$^a$ 
Takumi~Nagayama,$^b$ 
Kousuke~Suematsu,$^a$ 
Koichiro~Sugiyama,$^a$
Asao~Habe,$^c$ 
Mareki~Honma,$^d$ 
Noriyuki~Kawaguchi,$^d$ 
Hideyuki~Kobayashi,$^d$ 
Yasuhiro~Koyama,$^e$ 
Yasuhiro~Murata,$^f$ 
Toshihiro~Omodaka,$^b$ 
Kazuo~Sorai,$^c$ 
Hiroshi~Sudou,$^g$ 
Hiroshi~Takaba,$^g$ 
Kazuhiro~Takashima,$^h$ and 
Ken-ichi~Wakamatsu$^g$ \\ 
\llap{$^a$}Yamaguchi University, Japan\\
\llap{$^b$}Kagoshima University, Japan\\
\llap{$^c$}Hokkaido University, Japan\\
\llap{$^d$}VERA/National Astronomical Observatory, Japan\\
\llap{$^e$}National Institute of Information and Communications Technology, Japan\\
\llap{$^f$}Japan Aerospace Exploration Agency, Japan\\
\llap{$^g$}Gifu University, Japan\\
\llap{$^h$}Geographical Survey Institute, Japan\\
}

\abstract{
We present the basic features and the activities of Japanese VLBI network~(JVN), a newly-established VLBI network with baselines ranging from 50 to 2560~km spreading across the Japanese islands, and capable of observing at 6.7, 8.4, and 22~GHz.  We show a number of results of JVN observations: 8.4-GHz continuum images of a Giga-hertz Peaked Spectrum~(GPS) source and radio-loud Narrow-Line Seyfert~1 galaxies~(NLS1s), the spatial and velocity structures of water masers in NML~Cygni as well as methanol masers in Cep~A, and demonstrative observations with the bigradient phase referencing.
}

\FullConference{8th European VLBI Network Symposium\\
		 September 26-29, 2006\\
		 Toru\'n, Poland}

\begin{document}

\section{Introduction}
The activity of VLBI in Japan has been rapidly growing in this decade.  Japan is now one of the most highly populated area in the world in terms of VLBI antennas.  Following the idea to connect all Japanese radio telescopes (including the telescopes of VERA; e.g., \cite{Honma}) into a VLBI-imaging array, which is expected to collaborate with the Korean VLBI Network~(KVN; e.g., \cite{Shon}), other VLBI telescopes in the Asia-Pacific region, and the VSOP-2 (e.g., \cite{Hirabayashi}) as a ground network, Japanese VLBI Network~(JVN; Fujisawa et~al. in~prep.) was established.

\section{Japanese VLBI Network}
\begin{figure}[h]
\begin{center}
\includegraphics[width=0.66\textwidth]{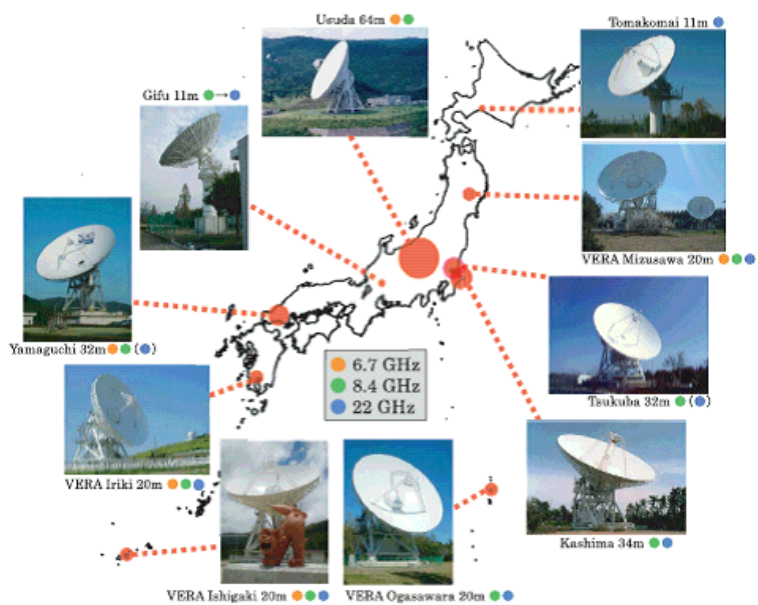}
\caption{Japanese VLBI Network.}
\label{figure1}
\end{center}
\end{figure}

JVN produced the first fringes in 2004 and the astronomical research has been conducted since 2005.  The array consists of ten antennas (Fig.~\ref{figure1}) that are owned and operated by four research institutes (National Astronomical Observatory of Japan~(NAOJ), Japan Aerospace Exploration Agency~(JAXA), National Institute of Information and Communications Technology~(NICT), and Geographical Survey Institute~(GSI)) and four universities (Hokkaido University, Gifu University, Yamaguchi University, and Kagoshima University).  These antennas form baselines in the range 50--2560~km across the Japanese islands and provide very dense $u$--$v$~coverage.  Three observing bands --- 6.7, 8.4, and 22~GHz --- are now available.  K4/VSOP-terminal system is currently used as a digital back-end with magnetic tapes at 128~Mbps.  Correlation processing is performed with the Mitaka FX correlator at NAOJ.  The sub-array of five telescopes -- Usuda~(64~m), Kashima~(34~m), Tsukuba~(32~m), Yamaguchi~(32~m), and Gifu~(11~m) -- are also connected with optical fibres at 2.4~Gbps.  This sub-array has already succeeded in providing real-time fringes on its ten baselines.  At the present stage, observations both with JVN and the fibre-connected sub-array are carried out based on the proposals put in by the Japanese VLBI community.

\section{Several observational results}
\begin{figure}[h]
\begin{center}
\includegraphics[width=0.63\textwidth]{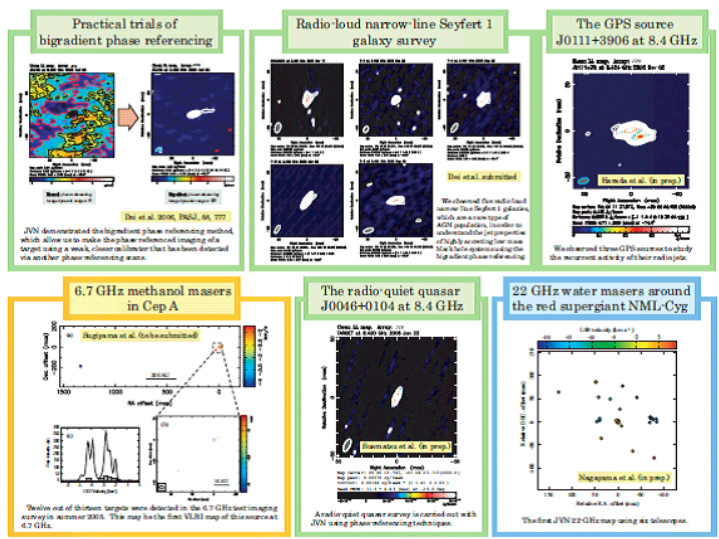}
\caption{A few scientific results obtained with JVN.}
\label{figure2}
\end{center}
\end{figure}

JVN has carried out scientific observations for more than 400~hours during this 1.5~years since its operations started on May 2005.  We have observing sessions throughout the year, $\sim3$~day/month on average.  The number of scientific results from JVN steadily increases at all three bands (Fig.~\ref{figure2}).  JVN has demonstrated the capability of Bigradient Phase Referencing~(BPR) method, which dramatically improves the phase-reference quality using a weak (i.e., fringe-undetectable) calibrator very close to a target \cite{Doi_etal.2006a}.  Using BPR at 8.4~GHz, we detected radio-quiet quasars~(RQQs; Suematsu et~al. in~prep.) and radio-loud narrow-line Seyfert~1 galaxies~(NLS1s) \cite{Doi_etal.2006b}, which had been rarely detected with the VLBI since they belong to AGN classes being very weak radio sources.  The giga-hertz peaked spectrum~(GPS) source J0111+3906 and other GPS sources were observed at 8.4~GHz in order to study the recurrent activity in their radio jets (Harada et~al. in~prep.).  The spatial and velocity structures of water masers in NML~Cygni (Nagayama et~al. in~prep.), methanol masers in Cep~A (Sugiyama et~al. to~be submitted), and many other cosmic maser sources have recently been obtained with JVN at 22 and 6.7~GHz.

\end{document}